\documentclass[apj]{emulateapj-rtx4}
\usepackage{color,ulem}
\usepackage{graphicx}
\usepackage{natbib}

\newcommand{\y}{\color{blue}}

\newcommand{\arXiv}{arXiv:}
\newcommand{\EPJWC}{EPJWC}
\newcommand{\Sci}{Science}
\newcommand{\PNAS}{PNAS}
\newcommand{\AN}{AN}

\newcommand{\g}{\, \rm g}
\newcommand{\cm}{\, \rm cm}

\begin{document}

\title{Warm Jupiters are less
  lonely than hot Jupiters: close neighbours}
\author
{
Chelsea Huang\altaffilmark{1,2},  Yanqin Wu\altaffilmark{3}, Amaury H.M.J. Triaud\altaffilmark{1,3,4}
}
\altaffiltext{1}{Centre for Planetary Sciences, University of Toronto at Scarborough, 1265 Military Trail, Toronto, ON, M1C 1A4, Canada}
\altaffiltext{2}{Dunlap Institute,University of Toronto, Toronto, ON, M5S 3H4, Canada}
\altaffiltext{3}{Department of Astronomy \& Astrophysics, University of Toronto, Toronto, ON, M5S 3H4, Canada}
\altaffiltext{4}{Institute of Astronomy, University of Cambridge, Madingley Road, Cambridge, CB3 0HA, United Kingdom}

\begin{abstract}

Exploiting the {\it Kepler} transit data, we uncover a dramatic
distinction in the prevalence of sub-Jovian companions, between 
systems that contain hot Jupiters (periods inward of 10 days) and 
those that host warm Jupiters (periods between 10 and 200 days). 
Hot Jupiters, with the singular exception of WASP-47b, 
do not have any detectable inner or outer planetary companions 
(with periods inward of 50 days and sizes down to $2 R_{\rm Earth}$). 
Restricting ourselves to inner companions, our limits reach down 
to $1 R_{\rm Earth}$. In stark contrast, half of the warm 
Jupiters are closely flanked by small companions. Statistically, the
companion fractions for hot and warm Jupiters are mutually exclusive, 
particularly in regard to inner companions.  \\

The high companion fraction of warm Jupiters also yields clues 
to their formation. The warm Jupiters that have close-by siblings 
should have low orbital eccentricities and low mutual inclinations. 
The orbital configurations of these systems are reminiscent of those 
of the low-mass, close-in planetary systems abundantly discovered by the 
{\it Kepler} mission. This, and other arguments, lead us to propose 
that these warm Jupiters are formed {\it in-situ}. There are 
indications that there may be a second population of warm 
Jupiters with different characteristics.
In this picture, WASP-47b could be regarded as the extending tail 
of the {\it in-situ} warm Jupiters into the hot Jupiter region, 
and does not represent the generic formation route for hot Jupiters.
\end{abstract}


\section{Forewords}

The origin of hot Jupiters (HJs, period inward of $\sim 10$ days) has
remained an unsolved issue. Although multiple scenarios have been
proposed (disk migration, planet scattering, secular migration, etc.),
none seem capable of satisfying all observational constraints. The
recent discovery of two low-mass planetary companions
\citep{Becker:2015} close to the hot Jupiter WASP-47b
\citep{Hellier:2012} further obfuscates the picture. Motivated by the 
large population of low mass, closely-packed planets at small
distances away from their host stars
\citep{Mayor:2011,Howard:2012,Borucki:2011,Lissauer:2011}, and by the
realization that some of them could have accumulated enough mass to
undergo run-away gas accretion \citep{LeeChiang:2014},
\citet{Boley:2015,Batygin:2015} argue that WASP-47b, and possibly all
hot Jupiters, were formed {\it in-situ}, instead of somehow 
transported inward. Only a tiny fraction of super-Earths need follow
this path to be able to match the occurrence rate of hot Jupiters.

While this seems a reasonable proposal for WASP-47b, could it 
explain the majority of hot Jupiters? To answer this, we focus on the
following issue: is WASP-47b a generic hot Jupiter in terms of
co-habiting with other planets? Currently, this question is
best addressed by exploiting the {\it Kepler} data to look for
small transiting bodies in systems hosting (either confirmed or
candidate) hot Jupiters. If we find that WASP-47b is truly unique
among all hot Jupiters, it may suggest that the formation of hot
Jupiters can have multiple pathways, with a minority being formed {\it
  in-situ}.

There is a second goal to our paper: understanding the warm Jupiters (WJs). 
By this term we refer specifically to those giant planets orbiting between 10 days and 
200 days in period. Unlike the hot Jupiters (inward of 10 days), they are 
too far out to have experienced little if any tidal circularization and 
therefore may be difficult to migrate inward by mechanisms that invoke 
high-eccentricity excitation. On the other hand, they live inward of the 
sharp rise of giant planets outside $\sim$ 1AU -- in fact, the period 
range of warm Jupiters corresponds to the so-called 'period-valley', 
the observed dip in occupation in-between the hot Jupiters and cold Jupiters 
\citep[e.g.,][]{Mayor:2011,Wright:2012,Santerne:2015}. In contrast with hot 
Jupiters, no theories have been proposed to explain the existence of 
this class of objects. So in this paper, we hope to gain some insights 
by studying their companion rates.


There have been multiple past claims that hot Jupiters lack sub-Jovian 
(and Jovian) close companions, by using the Radial Velocity data 
\citep{Wright:2009}, by inferring from (the lack of) transit timing 
variations in these objects 
\citep{Steffen:2005,Gibson:2009,Latham:2011,Steffen:2012}, and by 
searching for other transiting companions in the same systems \citep{Steffen:2012}.
The last study, in particular, is the closest to our work in spirit. 
Using preliminary candidates resulting from the first four months 
of the {\it Kepler} Mission (63 HJs and 31 WJs, defined differently 
from here), \citet{Steffen:2012} found a difference between the two 
populations: while none of the HJs have any transiting companions, 
five of the 31 WJ candidates do. These led them to suggest that 
HJs and WJs may be formed differently. However, due to the limitation 
of the short baseline of early stage {\it Kepler} data, and their 
crude criteria for candidate selection, their companion fractions 
are largely uncertain. For instance, among the five WJ candidates 
that were claimed to have companions: Kepler-18d, with a radius of 
$0.6 R_J$, is actually a hot Neptune \citep{Lithwick:2012}; KOI-190.01 
is a diluted eclipsing binary \citep{Santerne:2012}; and KOI-1300.01 is 
an eccentric eclipsing binary \citep{Ofir:2012}. This highlights 
the possible confusion ensued when selecting candidates based 
on early {\it Kepler} light-curves. Fortunately, four years down 
the road, we not only have the full four-year {\it Kepler} dataset 
at our disposal, we also have a large number of confirmed Jovian 
planets to inform us on the selection of candidates. We use 
both of these to our advantage and revisit the issue of 
companion fractions for close-in Jovian planets. 
        
Our paper proceeds as follows. We first describe how we select 
our samples of hot and warm Jupiters from the {\it Kepler} data 
and how we evaluate our selection completeness in 
Section \S \ref{sec:data}; the companion fractions of the two 
population are estimated and presented in 
Section \S\ref{sec:result}. We discuss the implication of our result 
in Section \S \ref{sec:discussion} on the formation paths of hot 
Jupiters and warm Jupiters.

\begin{figure}[h]
\centering
\includegraphics[width=0.5\textwidth]{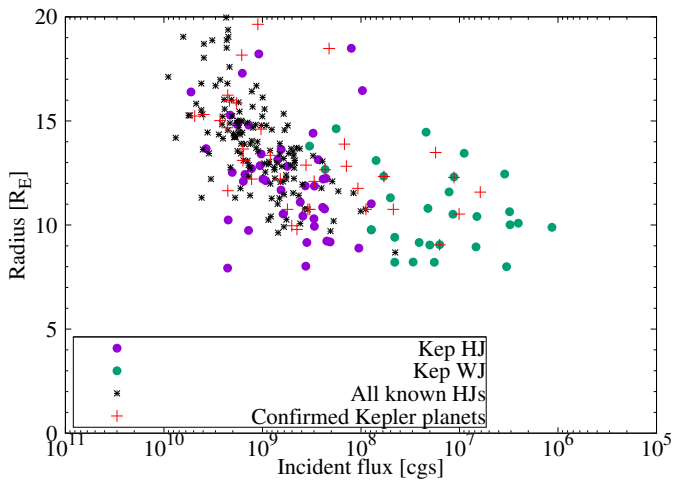}
\caption{The radius of our selected planets versus the incident irradiations 
    they receive, compared with all the known transiting hot Jupiters
  (black *) and confirmed {\it Kepler} giant planets (red crosses). We
  separated our sample into two groups, the hot Jupiters ($P<10 day$) are in
  purple dots, and the ones with longer orbital periods (warm Jupiters) in green
  dots. }
\label{fig:fluxradius}
\end{figure}

\section{Companion Fractions in {\it Kepler}}\label{sec:data}

\subsection{Sample selection}

We aim to construct a clean and complete giant planets sample 
with period range from 0.5 to 200 days from the
full set of {\it Kepler} data. We select the planet candidates 
with shorter period range (0.5 to 34 days) 
from the Kepler Object of Interests (\anchor{
http://exoplanetarchive.ipac.caltech.edu/cgi-bin/TblView/nph-tblView?app=ExoTbls\&config=cumulative\&constraint=koi\_pdisposition+like+\%27CANDIDATE\%27|}{cumulative table}
), 
and then compensate the list of longer period planets from existing 
literatures studying {\it Kepler} giant planets. We ensure a uniform selection 
of all candidates by making the same selection threshold as presented 
below on planet candidates from both origins. 

We start with all the candidates with radii between 8-20
$R_{\rm E}$. We then restrict ourselves to those
around stars with more reliable stellar parameters in the Kepler Input
Catalog (with stellar effective temperatures between 4500 and
6500K). To reduce the false positive rate in the sample, we also
require each of the planetary candidate to have a fitted impact
parameter smaller than 0.9, and a fitted stellar density from transit
parameters between 0.2$\g\,\cm^{-3}$ and 5$\g\,\cm^{-3}$
\citep{Seager:2003}.  We further remove candidates that have
detectable secondary eclipses, or ellipsoidal variations, or centroid
shifts during transit events.

The completeness and robustness of the {\it Kepler} candidates
decrease for giant planets with longer orbital periods. A few 
teams released their own catalogs of long period planetary 
candidates discovered within the {\it Kepler} data
\citep[i.e.]{Wang:2013}. In particular, \citet{Dawson:2015} selected 
31 KOIs with orbital periods longer than 34 days by using the combined 
catalog from all available sources (we refer to their Appendix A for 
their selection details).

We include the 22 candidates from \citet{Dawson:2015} that have
orbital periods between 34 day and 200 day to complement our above
selection, but reject five planetary
candidates from that sample because of their
V-shaped transits and/or anomalously large radius \citep[also see
``exceptional candidate treatment" section of][]{Dawson:2015}. These
warm Jupiters matched our selection criterion for the KOIs with
shorter orbital periods. Therefore combining both samples preserves a
uniform selection. We refer to our sample as the {\it Kepler} giant
planet sample in the text hereafter and do not specifically
distinguish in the text between the concept of ``planets" and ``planet
candidates".

We present our final sample in Tables
\ref{table:hjplanet} \& \ref{table:wjplanet}. This sample
includes all of the 40 confirmed giant planets from the
catalog summarized by \citet{Santerne:2015} matching 
our designed planet period and stellar properties range. 
With our definitions of hot Jupiters and warm Jupiters, 
our sample includes 45 HJs (28 confirmed), and 27 WJs (12 confirmed). 
It does not, however, include the hot Jupiter WASP-47b, since it is 
not observed by the {\it Kepler} main mission. 

In Fig.  \ref{fig:fluxradius}, we compare the sizes of these objects
with known Jupiters, as a function of their incident irradiations. 
They appear to fall into similar region, suggesting that 
the fraction of false positives in our sample is
small. More quantitatively, \citet{Morton:2012} developed a
statistical frame work to quantify false positive 
probabilities of {\it Kepler} candidates. Judging by
his values for the unconfirmed objects in our 
sample (Tables \ref{table:hjplanet} \& \ref{table:wjplanet}), 
the majority have negligible probabilities to be false positives.

\subsection{Search and completeness of small-sized planets}

We seek additional companions in systems hosting our
selection of giant planets. To address the possibility that the giant
planet transit signals would influence the detrending process and 
the transit search algorithm, we first removed all of their 
transits from the {\it Kepler} raw (simple aperture, SAP) light
curves. For each expected giant planet transit, this is done by 
creating a data gap in the light curve with a width $1.1\times$ the 
fitted transit duration and centred on the transit epoch. We then
detrended the light curves following \citet{Huang:2013} and searched for
additional transit signals using a Box Least Square (BLS)
algorithm. Any BLS peaks detected with transit dip significances
(signal-to-noise of the transit signal) higher than 10 are investigated to 
check if they are due to planet signals. 

\begin{figure}[h]
\centering
\includegraphics[width=0.5\textwidth]{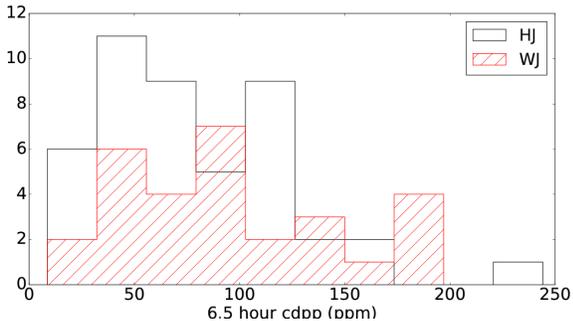}
\caption{Histogram of the 6.5 hour
  CDPP for the light curves of the host stars in our {\it Kepler}
  giant planet sample.  Those with hot Jupiters are shown in black,
  while those with warm Jupiters are shown in
  red. The two populations have similar noise properties.} 
\label{fig:cdpphist}
\end{figure}

Moreover, we can also calculate the completeness of detection 
for a planet of a given size and at a given period, based on the 
estimated signal-to-noise ratio,

\begin{equation}
{\rm SNR} = \frac{\delta}{\rm CDPP}\times \sqrt{N_{\rm transit}}\, .
\end{equation}
Here, $\delta$ is the expected transit depth for the planet, $N_{\rm
  transit}$ the number of transits in the entire {\it Kepler}
light curve, and CDPP the combined differential photometric precision
of the light curve over the transit duration time-scale.  
\citet{Fressin:2013}
reported that $99.9\%$ of the {\it Kepler} candidates with a
signal-to-noise ratio (SNR) larger than $10.1$ will be detected by the
{\it Kepler} pipeline.
When calculating the completeness, we define a planet as detectable 
if its SNR is greater than $10$. A
histogram of the 6.5-hour CDPP values for host stars in our sample 
is shown in Figure \ref{fig:cdpphist}. The host stars of HJs 
and WJs have similar noise properties in their light curves. So 
we expect similar detection completeness in these two groups.

\begin{figure*}
\centering
\includegraphics[width=0.8\textwidth,trim=50 0 140 40,clip=true]{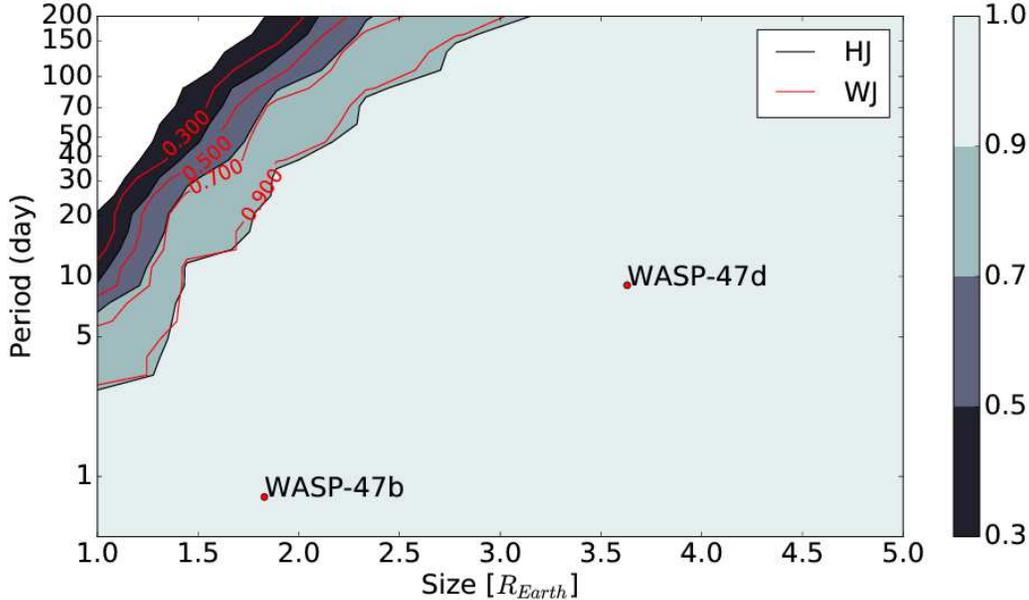}
\caption{The detection completeness 
    as a function of planet size (in Earth radius) and
    orbital period (in day) for our systems. The contours (and
    black lines) show the averaged completeness of all HJ systems,
    while the red lines show that for the WJ systems. They are
    similar. For planets larger than $2 R_{\rm Earth}$, we are $>90\%$
    complete out to $50$ days; while for those larger than $1 R_{\rm
      Earth}$, we are $>90\%$ complete out to $\sim 3$ days. The latter
    places stringent constraint on the inner companions of HJs.
    The interior and exterior companions of WASP-47c are noted here. 
    They are trivially detectable around all
  systems. }
\label{fig:completeness}
\end{figure*}

We report the detection completeness thus estimated in 
Figure \ref{fig:completeness}, averaged over the cohorts of HJ and WJ 
systems separately. As expected, they look rather similar. We further 
verify this estimation by performing a signal-injection-and-recovery
experiment. We inject into the HJ light curves transit signals by 
planets of orbit period around 10 days and of various sizes 
($2 R_{\rm Earth}, 1.5 R_{\rm Earth}$ and $1 R_{\rm
  Earth}$). We obtain recovery rates of 100$\%$, 93$\%$ and 51$\%$ at
these sizes. These match the above estimates using light curve CDPPs.

Also shown on Fig. \ref{fig:completeness} are the positions of the 
two small companions of WASP-47b. If present, they should be 
trivially detected around any of the giant planets in our sample.

\subsection{Results} \label{sec:result}

\begin{figure}
\centering
\includegraphics[width=0.5\textwidth]{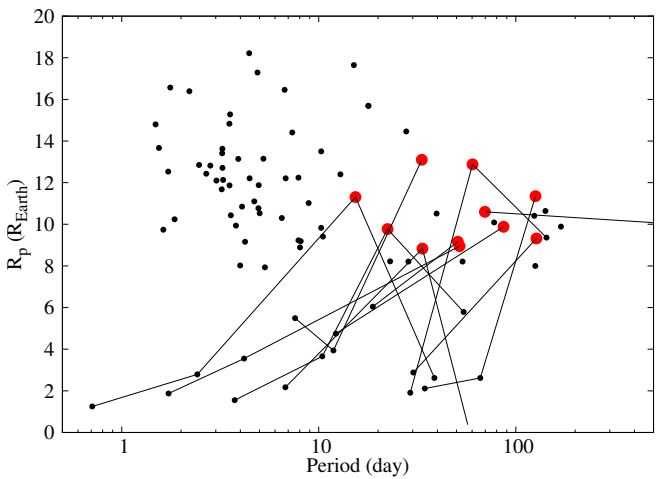}
\caption{Periods and radii for our sample of giant planets and their
  neighbours.  The giant
  planets that have companions are marked out in red and are
  connected to their companions by lines. Only warm Jupiters appear 
  to belong to multiple systems.
}
\label{fig:periodradius}
\end{figure}

\begin{figure}
\centering
\includegraphics[width=0.5\textwidth]{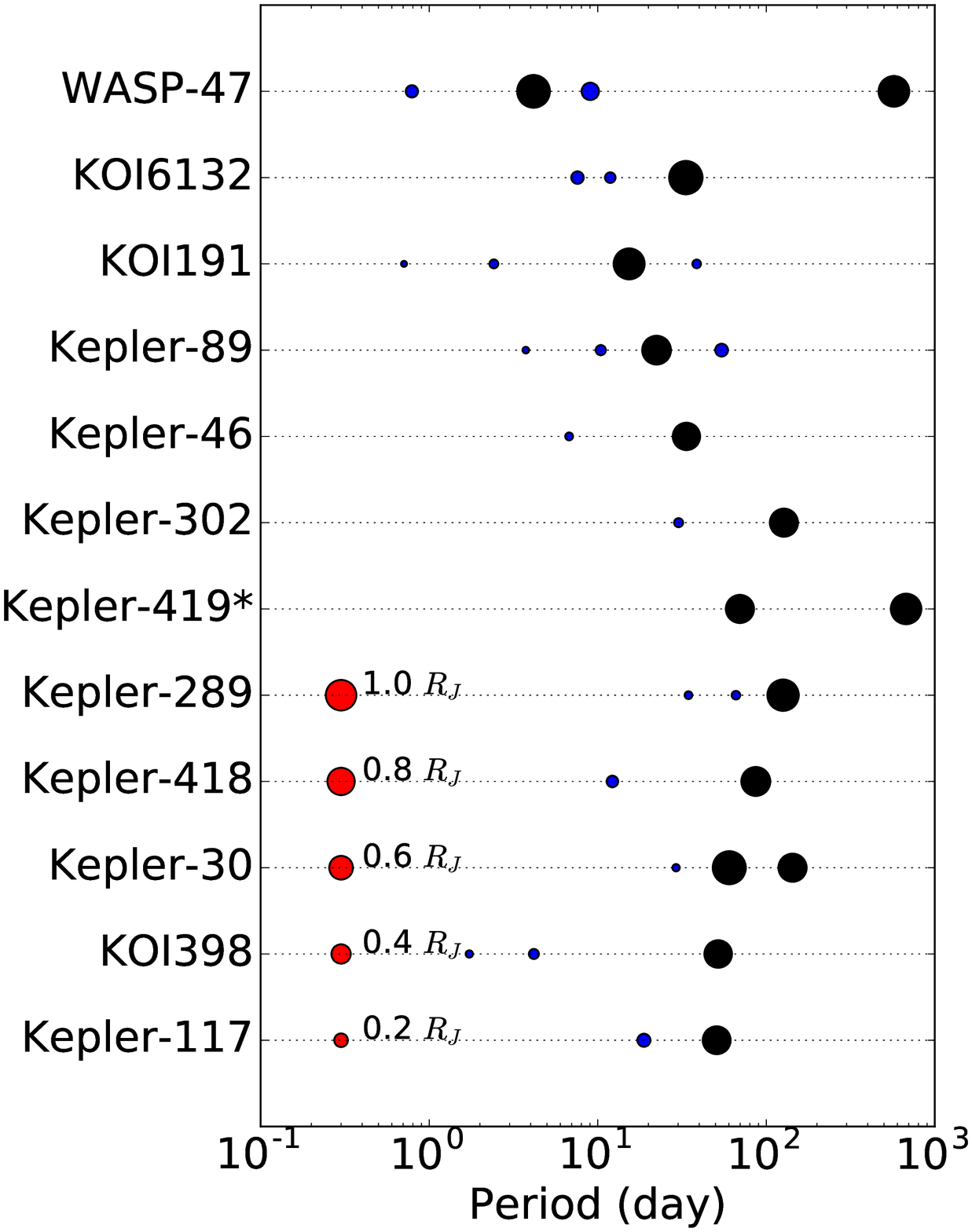}
\caption{Similar to Fig. \ref{fig:periodradius} but showing more
clearly the architectures of our multiple systems (around warm
Jupiters).  The planets are laid out in $\log P$, with the size of the
dot representing the size of the planet. The confirmed systems are noted
by their {\it Kepler} names. {\bf Kepler-419} system is counted as a single warm Jupiter 
system in our calculation, given the companion is far away.}
\label{fig:companions}
\end{figure}

We summarize our search results in Table \ref{tab:companions}. 
We find that, while none of the HJs have any transiting companions, 11 (out of 27) 
of the WJs do. All of these companions have also been reported by the 
{\it Kepler} candidate catalog. We display
the orbit architectures of these multiple systems in Figures \ref{fig:periodradius} \&
\ref{fig:companions}. 

\begin{table}
\centering
\begin{tabular}{ccccc}
\hline\hline
 group & total & $N_{\rm multi}$ & Inner & Outer\\
 \hline
 HJ & 45 & 0 & 0& 0\\
confirmed HJ & 28 & 0 & 0 & 0\\
 \hline
 WJ & 27 & 10 & 10 & 3\\
 confirmed WJ & 12 & 7 & 7 & 2 \\
 \hline
\end{tabular}
\caption{Summary of transiting companions for {\it Kepler} giant planets. }
\label{tab:companions}
\end{table}

Assuming all companions to the giant planets transit, we can estimate 
simply the companion fractions using results in Table 
\ref{tab:companions}. We focus on companions inward of $50$ 
days and larger than $2 R_{\rm Earth}$, since this population should 
be detected nearly completely ($>90\%$), and since 
most of the companions we find do fall in this range. We term these 
{\bf ``close''} companions.
We treat the problem as an estimation of the distribution of the event
success rate $p$ for a binomial distribution having observed $s$
successes in $n$ trials, for which each system is assumed to have
equal weight. We imply a conjugate prior (${\rm Beta}(0.5,0.5)$) on
$p$. 
The posterior distribution of $p$ can be expressed as ${\rm
  Beta}(s+0.5,n-s+0.5)$.  We thereby
obtain a multiplicity rate of hot Jupiters of
$0.52^{+5.0}_{-0.52} \%$, while that for warm Jupiters is $37.2^{+18.6}_{-16.3}\%$. 
Throughout this work, we present results in terms of 
the median of the distribution and their $90\%$ confidence interval.
  
If we only use the confirmed planet systems, the 
estimated multiplicity rate for hot Jupiters is ${\rm Beta}(0.5,28.5)=0.8^{+7.7}_{-0.8}\%$, 
and for warm Jupiters is ${\rm Beta}(8.5,4.5)=58.1^{+23.9}_{-26.9}$. 
This latter value, however, should be taken with caution as 
the confirmed systems are likely to bias towards multiple systems. 
As an example, they are usually prioritized during radial velocity follow-up observations. 
Other confirmation methods, such as transit timing variation \citep[i.e.]{Xie:2013}, 
or making use of low false positive rate of the Kepler multiply systems \citep[i.e.]{Rowe:2014}, 
also prefers to confirm multiple systems. 

 \begin{table*}
\centering
\caption{The probability for a giant planet to have at least one ``close'' companion 
in the system. \tablenotemark{a} \label{tab:finalrate}}
\begin{tabular}{ccccc}
\hline\hline
group & candidate fp & mutual inclination & HJ rate ($\%$) & WJ rate ($\%$)\\
\hline
1 & 0 & 0 & $0.5^{+5.0}_{-0.5}$ & $37.2^{+18.6}_{-16.3}$ \\
\hline
 2 & $10\%$ & $\sigma_{\mu}=1.8^o$ & $0.98^{+9.4}_{-0.98}$ & $58.1^{+31.7}_{-31.0}$\\
3 & $10\%$ & uniform & $8.4^{+46.4}_{-8.0}$ & -\\
\hline
4 & $50\%$ & 0 & $1.1^{+13.3}_{-1.1}$ &  $55.7^{+27.0}_{-31.1}$ \\
5 & $50\%$ & $\sigma_{\mu}=1.8^o$ & $1.6^{+19.5}_{-1.6}$ & $69.5^{+24.2}_{-30.8}$ \\
6 & $50\%$ & uniform & $9.1^{+45.8}_{-8.6}$ & -\\
\hline
7 & $100\%$ & 0 & $0.8^{+7.7}_{-0.8}$ & $58.1^{+23.9}_{-26.9}$\\
\hline
\tablenotetext{1}{For case 1,4 and 7, ``close" companion refer to those with period 
smaller than 50 days. For case 2,3,5 and 6, only the interior companion of the giant planets are considered.}
\end{tabular}
\end{table*}

\begin{figure}
\centering
\includegraphics[width=0.5\textwidth,trim=0 50 80 20,clip=true]{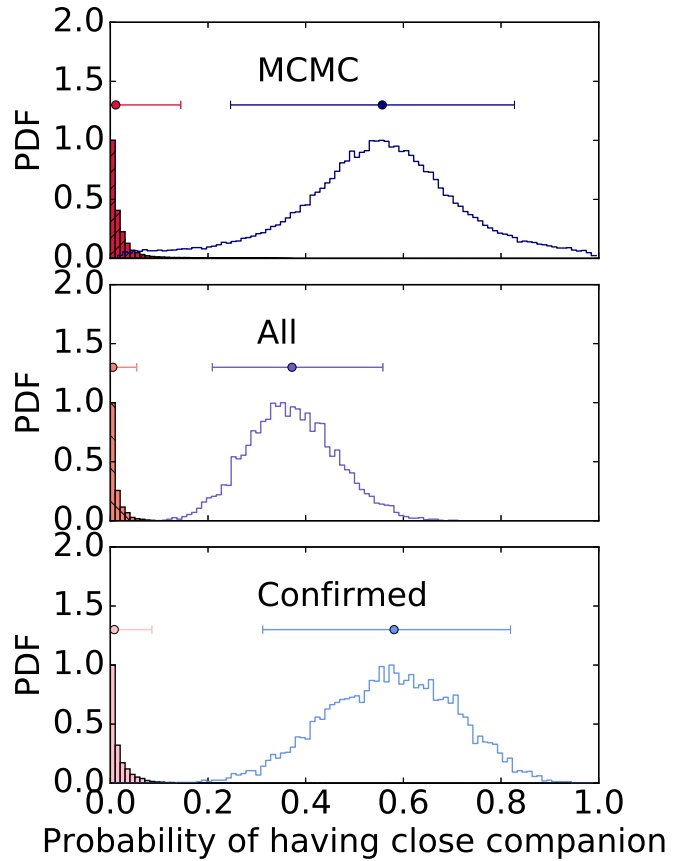}
\caption{The probability density distributions for
  giant planets to have ``close'' companions, defined as inward of
    $50$ days and larger than $2 R_{\rm Earth}$. Here we assume that
    all planets in a given system transit. The top panel
  compares the companion rates for hot Jupiters (blue) and warm
  Jupiters (red) in our standard Monte Carlo model 
    where the candidate false positive rate is assumed to be $50\%$.  
    Points with error-bars indicate the median of the
  distribution and its $90\%$ confidence interval. The middle panel
  assumes that all unconfirmed candidates are
  real planets ($fp = 0$), whereas the bottom panel assumes that
  they are all false positives ($fp = 100\%$) and only confirmed
    planets are included.  The two populations of giant planets
    are incompatible in all cases.
 }
\label{fig:rate}
\end{figure}

\begin{figure}
\centering
\includegraphics[width=0.5\textwidth,trim=5 20 80
10,clip=true]{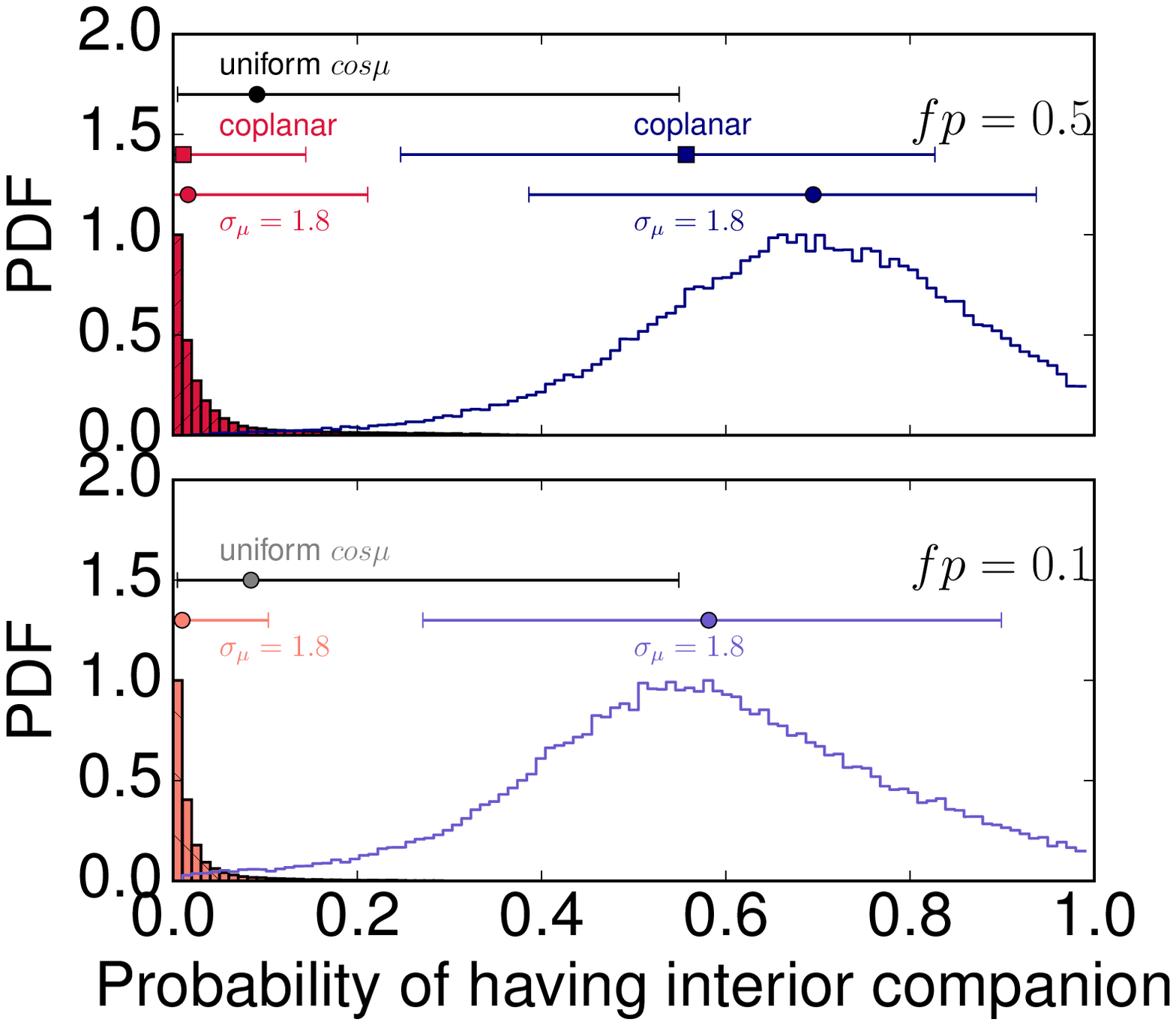}
\caption{Same as Fig. \ref{fig:rate} but demonstrating the impact of
    mutual inclinations, considering only {\em inner} companions. 
    The top panel compares the {\em inner} companion
  rates using our standard priors, with the
    histograms corresponding to a case where the mutual inclinations
    is a narrow Rayleigh distribution with a dispersion of
    $\sigma_{\mu} = 1.8^o$ -- compare these to the co-planar case.
  The extreme case, marked with ``uniform $cos\mu$'' for the HJ only,
    is one where we assume that the orbital planes for hot Jupiters and their inner
    companions are un-correlated. The bottom
  panel shows the same content but for a smaller false positive
  rate of $fp = 10\%$, a value that we deem more plausible for our
  sample.}
\label{fig:rate_interior}
\end{figure}

\subsubsection{Monte Carlo Simulation}

To have a more realistic estimation that takes into account of a
multitude of observational uncertainties, we perform a Monte Carlo
Markov Chain simulation with emcee \citep{Foreman-Mackey:2013} to
constrain the rate of multiplicity. 

For our standard MCMC model, we again assume all companions transit (a
flat system), and we adopt a false positive rate of $50\%$ for the
unconfirmed planets in our sample. We continue to focus on
  ``close'' companions for which our detection capacity is nearly
  complete.  We also account for the scatter of the Neptune
population into our sample. The details are laid out in Appendix
\ref{sec:stats} and the results are shown in the top panel of Figure
\ref{fig:rate}. We obtain that hot Jupiters have a multiplicity rate
of $1.1^{+13.3}_{-1.1}\%$, and warm Jupiters have a multiplicity rate
of $55.7^{+27.0}_{-31.1}\%$, largely unchanged from our previous
estimates.

In the following, we relax or alter some of the assumptions in our 
standard model. Results of all these experiments are summarized in 
Table \ref{tab:finalrate}.

\subsubsection{Impact of false positives}

Fig. \ref{fig:rate} compares the results when 
different false positive rates for the unconfirmed systems are 
adopted. The overall effect of a smaller false-positive rate is to 
reduce the companion fraction for WJs, since the confirmed WJs have 
a higher multiplicity than the unconfirmed ones.

We note that the false positive rate we adopt for our standard model 
($fp = 50\%$) is likely too pessimistic for our sample. 
This value is derived by \citet{Santerne:2015} for all giant 
candidates in the entire {\it Kepler} candidate sample, while we 
benefit from additional filtering based on follow-up measurements 
and additional vetting on the light curves. In fact, the individualized 
false-positive rates (Morton et al, 2016, in prep) for the unconfirmed 
objects in our sample are in general low. We regard an optimistic 
value of $fp =10\%$ as calculated from the above work to be 
more suitable for our WJ candidates. 
We will present more results for this false positive rate below.

\subsubsection{Impact of mutual inclination}

So far, we have assumed that the companions always
transit. \citet{Fabrycky:2014} found that the {\it Kepler} multiple
systems are indeed quite coplanar \citep[also
see][]{Figueira:2012,FangMargot:2012,Tremaine:2011}.  Assuming that
the mutual inclination ($\mu$) dispersion follows a Rayleigh
distribution, the inferred dispersion $\sigma_\mu$ typically lies in
the range 1.0$^{\circ}$-2.2$^{\circ}$. We investigate how this 
affects our results. 

We consider only the possibility that the giant planet systems to have a
non-transiting companion interior to the giant planet. 
As the transit probability is now a function of orbital period, 
we need to assume a period distribution for the companion.
The occurrence rate of sub-Jovian planets is roughly flat in the log 
period space, and falls off for short periods \citep{Howard:2012,Dong:2013}. 
We assume the companion occurrence rate is flat in $\log P$ for
WJs, with a inner cut-off period of 1 day, and a period power-law dependence 
for period for HJs, with a inner cut-off period of 0.5 day. We simulate 
the rate of a interior companion assuming the prior for the mutual inclination 
follows a Rayleigh distribution with $\sigma_{\mu}=1.8^{\circ}$ for both the HJs and
WJs. To explore the extremes, we also simulate the rate of interior
companion for HJs assuming the $\cos\mu$ distribution is uniform
between 0 and 1. The distribution of interior companion rate is
constrained as in Figure \ref{fig:rate_interior}, upper panel using a false positive rate of 
$50\%$. We show that assuming the same mutual inclination distribution, the median of
the multiplicity rate distribution of HJs and WJs both shift to a
higher value, but their $90\%$ confidence interval do not overlap with
each other. Even with the extreme case that the HJ companions have an
uniform cosine mutual inclination distribution, the $90\%$ confidence
interval of its distribution exclude the median of the WJ companion
rate distribution. The conclusion holds with a $10\%$ false positive 
rate (lower panel of Figure \ref{fig:rate_interior}). If we assume instead a 
uniform logP distribution for the HJ companions, similar to that of the WJ's, the
conclusion will be stronger since it allows less non-transiting
companions.

For interior companions of HJs, we are able to constrain the detection 
of planets with sizes larger than 1.5 $R_{\rm Earth}$ to be complete 
up to 10 days. However, this conclusion can be extended to 1$R_{\rm Earth}$ 
if we correct for detection completeness assuming $df/dlogR$ is constant 
between 1-3 $R_{\rm Earth}$ (as suggested by \citet{Petigura:2013}).
In reality, the majority of the HJs have orbital period smaller than 5 days, 
allowing us to have almost complete detection to 1 $R_{\rm Earth}$ regardless of 
the size distribution.

The rate of exterior companions is extremely sensitive to the distribution 
of mutual inclinations, as such it is difficult to obtain a good constraint. If we 
assume every giant planet with interior companion also have an exterior companion, 
and that the interior and exterior companions have the same mutual inclination 
distribution (a Rayleigh distribution with $\sigma_\mu=1.8^{\circ}$), the 
observed number of systems with exterior companion is expected to be a 
quarter of those with interior companions due to the transit probability 
and detection completeness. This is not in contradiction with the current 
observations. 

\section{Discussions}\label{sec:discussion}

Our main results (Fig. \ref{fig:rate} and \ref{fig:rate_interior}) 
that HJs have a companion fraction consistent with zero, while about
half of WJs have close neighbours, reveal much about
their respective formation mechanisms.

\subsection{Two populations of Warm Jupiters?}

Two previous works have suggested that warm Jupiters are not all the
same beasts.  Using RV data, \citet{Dawson:2013} suggested
that eccentric WJs tend to orbit around metal rich stars, while
planets around metal poor stars have predominately circular orbits.
Later, also based on RV data, \citet{Dong:2014} showed that more than
half of the eccentric WJs ($e > 0.4$) have distant, Jovian companions,
while the low-eccentricity WJs tend to be 'single'.  Taken together,
these suggest that more Jovian planets are produced around metal-rich
stars, and this somehow raises the eccentricities of WJs (via either
secular perturbations or planet-planet scatterings). 
Recently, \citet{Bryan:2016} confirmed the above finding with a larger sample. 
In addition, they reported that WJs have a lower occurance rate of 
distant giant companions (defined as with masses between 1-20 $M_J$, and at 
orbital distances of 5-20 AU),  compared to what they found regarding 
the HJs, suggesting that dynamic processes due to giant planetary companians 
are less important in the WJ systems.

Our study of the {\it Kepler} transit data adds a further dimension to
the picture. At least half of the WJs in our sample are flanked by
close, small companions.  Most likely, they are nearly coplanar with
these companions (to maximize transit probability), and are also
nearly circular (to prevent dynamical instability in such
closely packed systems). RV counter-parts to these objects have been
found, e.g., HIP 57274c \citep{Fischer:2012} which is a nearly
circular WJ with a small mass neighbour interior to its orbit, and
 GJ 876c \citep{Rivera:2010}, and 55 Cnc-b
\citep{Endl:55}. However, due to limits on RV precision, most of these
WJs will in general be observed as 'singles' on largely circular
orbits.\footnote{In other words, the transit technique is uniquely
  capable of revealing these small neighbours.}

How are the warm Jupiters formed? For those with low-mass neighbours,
we postulate that they are formed {\it in-situ}, i.e.,
they undergo run-away gas accretion locally, and that the abundant, 
closely packed, super earth popolation discovered by {\it Kepler} 
are the cores enabling the accretion.  This hypothesis is supported by the
following arguments.

\begin{itemize}


\item Small planets (sub-Jovian) are common around stars, and their
  frequency rises outward steeply at just the inner edge of our WJ
  zone \citep{Howard:2012}, and it remains flat in logarithmic period
  beyond this cut-off \citep{Petigura:2013b,Silburt:2015}. 

\item Measured masses for some of these small planets are available
  \citep[e.g.][]{Weiss:2013,Lithwick:2014}, with some reaching beyond
  $10 M_{\rm Earth}$ in solid mass.

\item Such massive cores can retain a heavy enough atmosphere and
  undergo run-away gas accretion, at the observed locations for these
  WJs \citep[e.g.][]{Rafikov:2006}. Recent theoretical studies of
  accretion of such an envelope \citep{LeeChiang:2014,Batygin:2015}
  further argue that cooling can be sufficiently fast to allow {\it
    in-situ} formation of giant planets at these distances.

\item As the WJ grows in mass, it can keep the planetary configuration
  largely intact.\footnote{It will be able to destabilize some of the
    closest neighbours and ingest them in the process.} As a result,
  we observe them today still flanked by small neighbours.

\end{itemize}

Alternative theories such as disk migration,
whereby these WJs are formed at larger distances and migrated inward
by interaction with the protoplanetary disk, and
high-eccentricity migration (such as Kozai migration, planet-planet scattering, etc), 
face the challenge of explaining the presence of small neighbours. 

What about the other WJs, the ones that have no transiting companions?
Studies mentioned above argue that these may be a distinct
  population. However, it needs to be firmly
established that these WJs indeed always have distant Jovian
companions, and/or are largely eccentric, and/or only reside around
metal-rich stars.  At
the moment, their origin remains an enigma.

Assuming that the lonely WJs in our sample are from a distinct
  population, we can estimate the
relative proportions of these two groups of WJs.  Among the WJs, 10
out of 27 are in multiple systems, or $37^{+19}_{-16}$\% of
WJs. However, this value can rise to $58^{+31}_{-31}\%$ if the mutual
inclination dispersion of the system follows a Rayleigh distribution
peaked at $1.8 \deg$. In other words,
at least half of all WJs are formed {\it in situ}.

\subsection{Formation of Hot Jupiters}

We were initially motivated by the discovery of close companions
around the hot Jupiter WASP-47b \citep{Becker:2015}. After having
established the general patten among hot and warm Jupiters, we return
to reflect upon the existence of this particular system.

WASP-47b was first identified by ground-based photometry, and 
then verified with radial velocity measurements
\citep{Hellier:2012}. Thanks to this detection, it was proposed as a
target for the K2 mission. In the meantime, the radial-velocity
monitoring of the system unveiled a second, long period gas giant
\citep{NeveuVanMalle:2015}.  \citet{Becker:2015} analyzed the K2 data
and announced the presence of two super-Earths, one inner to the hot
Jupiter, and one between both gas giants. The two low-mass companions
of WASP-47b were discovered in a targeted observation, so it is
difficult to ascertain their statistical importance. However, we can
set an upper limit on the frequency of such objects, based on the
absence of neighbours in our hot Jupiter sample. This is $1.0^{+9.5}
_{-1.0}\%$ among all hot Jupiters, assuming a candidate false positive
rate of $10\%$ and a mutual inclination dispersion of $1.8 \deg$
(Table \ref{tab:finalrate}).

This leads us to believe that WASP-47b is a rather unique system among
hot Jupiters, themselves rare systems. Based on our
discussion above, we further speculate that since more than half of
WJs may be formed {\it in-situ}, this path may found its way into the
period domain of HJs. Or, WASP-47b is also formed {\it in-situ}
\citep{Boley:2015,Batygin:2015}, and is a tail of the {\it in-situ}
WJs.  Given the relative numbers of hot Jupiters (one, WASP-47b) and
warm Jupiters (7 out of the confirmed sample of 12) in multiple 
systems, and the lower transit probability of WJs, this would 
suggest that the {\it in-situ} process is increasingly difficult 
for closer-in planets, with WASP-47b being the hottest 
representative of the warm Jupiter population.


Where do majority of the hot Jupiters come from then? We 
turn to the observations for answer.
Many of the dynamical migration processes that are invoked to
  explain hot Jupiters invariably produce the so-called 3-day pile
up, an excess of HJs near 3-day orbital period. Such a pile-up,
originally discovered among RV population \citep{Wright:2009}, but
then found to be absent in the {\it Kepler} transit data
\citep{Howard:2012}, and then re-discovered recently through careful
RV falsification of the transit candidates \citet{Santerne:2015}, now
seems here to stay. Hot Jupiters went through dynamical migration also 
led to destructions of any inner planetary systems naturally \citep{Mustill:2015}. 
This lends support to the hypothesis that hot
Jupiters are products of violent dynamical processes.

\section{Concluding words}

When we set out on this project, we wondered whether WASP-47b is a
common-place hot Jupiter in terms of small neighbours.  By examining
the Kepler hot Jupiters, we found that systems such as WASP-47b are
rare among the hot Jupiters -- among our statistical sample of 45 (28
confirmed) hot Jupiters, none show small companions, either inner or
outer.

In contrast, we found that among our warm Jupiter sample, half or more
have nearly coplanar, small companions.  Most of
  these are inner companions, and given the fact that exterior
  companions are less likely to transit, the data give the
  impression that they are at least as common as the inner ones (Table
  \ref{tab:companions}).

So not only do hot Jupiters and warm Jupiters appear to be separated
in their period distributions \citep{Santerne:2015}, they are also
distinct in their respective fractions of close neighbours.

Motivated by this discovery, and by recent theoretical progress in
understanding gas accretion, we propose that a significant fraction of
warm Jupiters are formed {\it in-situ}. 
The prevalence of multiple low-mass planets in close proximity to one
another and to the star can, in a fraction of the cases, permit some
of the planets to accrete enough envelope and to trigger run-away
growth. This process can operate in the warm Jupiter locale, but 
appears to become increasingly difficult towards the hot Jupiter region,
explaining the rarity of systems like WASP-47b.

We outline future directions below ,
\begin{itemize}

\item The true neighbour fractions of warm Jupiters is sensitive to
  the inclination dispersion in the system, as well as the false
  positive rate of our warm Jupiter sample (Table
  \ref{tab:finalrate}). Both could be improved. Specifically, further
  photometric and Radial Velocity monitoring of the confirmed warm
  Jupiters discovered by ground based transit surveys, such as
  HAT-P-15b \citep{Kovacs:2012}, HATS-17b \citep{Brahm:2015}, are
  likely to reveal the presence of companions, if
  current statistics holds.

\item Masses for these close companions are interesting. They may
yield the critical mass above which runaway gas accretion occurs.

\item Multiple lines of evidences suggest that there may be a second
  population of warm Jupiters. But data are inconclusive at the
  moment. Studies of warm Jupiter eccentricity distribution, host star
  metallicity distribution, presence of giant planet companions,
  etc. may help answer whether there is a second pathway to form warm
  Jupiters.


\item The {\it in-situ} warm Jupiters accrete their gas inward of 
    the ice-line. As such, their envelope water content should 
    not be depleted by condensation. This may contrast with those 
    that accrete their gas beyond the iceline (barring core erosion 
    in these bodies) and may be testable by transmission spectroscopy.


\item Using transit data, we can exclude inner neighbours of hot
  Jupiters, but the constraint is less stringent for the outer
  neighbours, especially if they may be highly misaligned
  \citep{Batygin:2015}.  Radial velocity study is necessary.

\item If outer neighbours are indeed also absent among hot Jupiters,
  theoretical study is needed to explain the rarity of {\it in-situ}
  formation for hot Jupiters, vs. that of warm Jupiters.

\end{itemize}

\section*{Acknowledgements}
The Dunlap Institute is funded through an endowment established 
by the David Dunlap family and the University of Toronto.
References to exoplanetary systems were obtained through the use of
the paper repositories,
\anchor{http://adsabs.harvard.edu/abstract\_service.html}{ADS} and
\anchor{http://arxiv.org/archive/astro-ph}{arXiv}, but also through
frequent visits to the \anchor{http://exoplanet.eu}{exoplanet.eu}
\citep{Schneider:2011} and
\anchor{http://exoplanets.org}{exoplanets.org} \citep{Wright:2011}
websites.  This paper includes data collected by the Kepler
mission. Funding for the Kepler mission is provided by the NASA
Science Mission directorate. Some of the data presented in this pa-
per were obtained from the Multimission Archive at the Space Telescope
Science Institute (MAST). STScI is operated by the Association of
Universities for Research in Astronomy, Inc., under NASA contract
NAS5-26555. Support for MAST for non-HST data is provided by the NASA
Office of Space Science via grant NNX09AF08G and by other grants and
contracts. This research has made use of the NASA Exoplanet Archive,
which is operated by the California Institute of Technology, under
contract with the National Aeronautics and Space Administration under
the Exoplanet Exploration Program.

\appendix

\section{Statistical Frame Work}
\label{sec:stats}

The observed giant planet multiplicity systems $N_{\rm obs,m}$ can 
arise from four origins: the confirmed multiple planet 
systems $N_{\rm obs,p,m}$, candidate multiple planet 
systems $N_{\rm obs,gc,m}$ with at least a giant planet, candidate 
multiple planet systems with the most massive planet being Neptune 
nature $N_{\rm obs,nc,m}$, and false positive multiple systems due to 
binary contamination $N_{obs,s,m}$. We assume the 
contribution to the multiple system with a giant planet from 
false positives ($N_{\rm obs,s,m}$) to be negligible \citep{Lissauer:2014}.  

\begin{equation}
N_{\rm obs,m} = N_{\rm obs,p,m} + N_{\rm obs,gc,m} + N_{\rm obs,nc,m}. 
\end{equation}

The individual terms above can be expressed as the following: 

\begin{equation}
N_{\rm obs,p,m} = N_{\rm p}f(g,m)f(d), 
\end{equation}
in which, $N_{\rm p}$ is the total number of confirmed system we have, $f(g,m)$ is 
the rate of giant planet system having a close-in companion, $f(d)$ is the 
probability for this companion to be detected. 

The number of multiple systems originated from ``Jupiters" and ``Neptunes" can be 
estimated as:
\begin{equation}
N_{\rm obs,gc,m} = N_{\rm pc}f(g)f(g,m)f(d),
\end{equation}
and 
\begin{equation}
N_{\rm obs,nc,m} = N_{\rm pc}(1-f(g))f(n,m)f(d),
\end{equation}
in which $f(g)$ is the fraction of giant planets in the our selected radius 
range, $f(g,m)$ and $f(n,m)$ is the probability a Jupiter/Neptune size planet 
having a close-in companion.  
$N_{\rm pc}$ is the number of planetary candidates that has planet nature, 
\begin{equation}
N_{\rm pc} = (N-N_{\rm p})\times(1-fp), 
\end{equation}
with $fp$ denote the false positive rate of a unconfirmed KOI.

The problem can be summarized as a Bayesian problem, in which $N_m$ is 
observable, and $f(g,m)$,$f(n,m)$,$f(FP)$,$f(g)$ as parameters we want to 
constrain given the data. We can do this by sampling the posterior space: 
\begin{equation}
p(\hat{f} | \hat{w_{\rm obs}}) = p(\hat{f})\times\,p(\hat{w_{\rm obs}} | \hat{f}).
\end{equation}
The likelihood $p(\hat{w_{\rm obs}} | \hat{f})$ can be expressed 
as $B(N_{\rm obs,m}-N_{\rm obs,nc,m} | f(g,m), N_p+N_{\rm pc,g})$, in which $B(s|x,N)$ indicate the probability of observing s success given N observations with success rate $x$.  

The priors $p({\hat{f}})$ are assumed (estimated) as below. 
\begin{itemize}
\item We use a conjugate prior for the multiplicity rate of giant planet $f(g,m)$. 

\item $f(g)$ is the probability of an unconfirmed planet to be a giant planet. We 
    use the posterior of the radius distribution of the candidate to decide the 
    probability of a candidate to have a radius smaller than 8$R_{\rm Earth}$. We 
    derived the average probability of a unconfirmed HJ to be a Neptune to be Beta(2.06,15.9), 
    and for warm jupiter is Beta(2.65,18.4).
 
\item We use the multiplicity rate for the super Neptunes ($4R_{\rm Earth}<R_p<8R_{\rm Earth}$) 
    within our designed stellar parameter and period range, in the entire KOI sample 
    (25 out of 140), as a prior for $f(n,m)$. We note that this rate is quite 
    uncertain, \citep{Mayor:2011} report the multiplicity rate for small planets 
    are $\sim70\%$ with an unclear statistic significance. Given the rate of $f(g)$ is 
    small, the choice of f(g) do not impact the final result significantly.

\item We use two sets of false positive rate. We first use the false positive rate 
    from \citep{Santerne:2015} to estimate the false positive rate of an unconfirmed 
    candidate $fp$. \citet{Santerne:2015} identified 46 false positives out of 100 KOIs 
    in the effective temperature and period range we use. For hot Jupiters, the FP rate 
    is $42.8\%$, and $49\%$ for warm Jupiters. This is lower than the overall false 
    positive rate reported by \citet{Santerne:2015}, since the authors found that the 
    false positive rate around stars with $T_{\rm eff}$ higher than 6500 K is generally 
    higher, which are not included in our sample. We apply a beta distribution prior for 
    the false positive rate. Due to our additional selection on our candidates, the actual 
    false positive rate can be even lower. We use the astrophysical false positive rate 
    estimated by Morton et al (in prep) to obtain a more optimistic value (10\%).

  \item We do not attempt to factor in the actual value of
    the detection completeness $f(d)$ for our estimation of the final
    rate since the size distribution of small planets have big
    uncertainties. We note that the detection is complete to 2$R_{\rm
      Earth}$ with orbital period less than 50 days for both HJs and
    WJs. While for interior companions, it is complete for HJs to
    1.5$R_{\rm Earth}$. We are able to prove that our conclusions are not changed if 
    extend this size limit to 1$R_{\rm Earth}$ 
    by assuming $df/dlogR$ is constant between 1-3 $R_{\rm Earth}$ as 
    suggested by \citet{Petigura:2013}. Our experiment found out 
    that counting smaller planets is equivalent
    to assume a more dispersed mutual inclination distribution for
    both HJs and WJs. Unless the occurrence rate of planet has a 
    steep rise towards small size, it does not impact the companion rate of 
    HJs much, while the companion rate of WJs may shifts to a higher value. 

\end{itemize}

To take into account the effect of mutual inclination angle dispersion, we compute an 
averaged transit probability for HJs and WJs separately on a mutual inclination grid 
with $\cos\mu$ from 0 to 1. We later on interpolate on this grid to obtain the transit 
probability at an arbitrary mutual inclination angle in our simulations. Since the 
inclination angle of the giant planet $i_g$ is usually loosely constrained by the transit 
fit, we assume that $\cos{i_g}$ has a uniform probability as long as the planet transit. We 
also integrate over the possible period range the companion could take. 
\begin{equation}
\bar{f}_{\rm tran}(\mu) = \frac{1}{N} \int f_{\rm tran}({\rm period},i_g,\mu) P({\rm period})P(i_g)\,{\rm d}{\rm period}\,{\rm d}i_g
\end{equation}

For the probability of the companion occur at a certain orbital period, we use the formalism 
from \citet{Howard:2012} for the HJs. 
\begin{equation}
{\rm d}f/{\rm d\,log} ({\rm period})\propto {\rm period}^{\beta}(1-e^{-({\rm period}/7.0 {\rm day})^{\gamma}}),
\end{equation}
in which, $\beta=0.27\pm0.27$, $\gamma=2.6\pm0.3$. For ${\rm period} <10 day$, this can be 
approximated with $f({\rm period})\propto {\rm period}^2$. We use the latter in our calculation. 
For WJs, we assume ${\rm df}/{\rm dlog}({\rm period})=C$ instead with a
cut-off inner period at 1 day. 

The likelihood expression can be revised as $B(N_m | f(g,m)f_{\rm tran}, N_p+N_{\rm pc,g})$ 
as a conditional Binomial problem. Here we are assuming that the covariance between $f(g,m)$ 
and $f_{\rm tran}$ are 0. 
 
We choose two type of priors for the mutual inclination of HJs, a Rayleigh distribution 
with $\sigma_\mu=1.8$ and a uniform distribution for $\cos\mu$ between 0 and 1. We only use 
the Rayleigh distribution prior for the WJs.

\defcitealias{Almenara:2015}{Al12}
\defcitealias{Bonomo:2015}{Bo15}
\defcitealias{Rowe:2014}{Ro14}
\defcitealias{Bruno:2015} {Br15}
\defcitealias{Nesvorny:2012}{Ne12}
\defcitealias{Weiss:2013}{We13}
\defcitealias{He:14}{He14}
\defcitealias{Steffen:2010}{St10}
\defcitealias{Cabrera:2013}{Ca14}
\defcitealias{Schmitt:2014}{Sc14}
\defcitealias{Sanchis:2013}{Sa12}
\defcitealias{Schmitt:2014b}{Sc14b}
\defcitealias{Dawson:2012}{Da12}
\defcitealias{Santerne:2015}{Sa12}
\defcitealias{Esteves:2015}{Es15}
\defcitealias{Shporer:2011}{Sh11}
\defcitealias{OD:06}{Od06}
\defcitealias{Southworth:2010}{So10}
\defcitealias{Lillo-box:2015}{Li15}
\defcitealias{Pal:2008}{Pa08}
\defcitealias{Van:13}{Va13}
\defcitealias{Buchhave:2011}{Bu11}
\defcitealias{Bonomo:2012}{Bo12}
\defcitealias{Deleuil:2014}{De14}
\defcitealias{Jenkins:2010}{Je10}
\defcitealias{Latham:2010}{La10}
\defcitealias{Gandolfi:2013}{Ga13}
\defcitealias{Santerne:2012}{Sa12}
\defcitealias{Faigler:2013}{Fa13}
\defcitealias{Endl:2014}{En14}
\defcitealias{Koch:2010}{Ko10}
\defcitealias{Endl:2011}{En11}


\begin{deluxetable*}{llccccl}
\tabletypesize{\footnotesize}
\tablecolumns{7}
\centering
\tablewidth{0pt}
\tablecaption{Kepler Hot Jupiter Systems \label{table:hjplanet}}
\tablehead{
\colhead{KIC}&\colhead{Period}&\colhead{$R_p$}&\colhead{$M_p$}&\colhead{$R_{\rm star}$}
& \colhead{fp rate{\tablenotemark{a}}} &\colhead{Other Name/Reference{\tablenotemark{b}}}  \\
\colhead{}&\colhead{[day]}&\colhead{$[R_{\rm
    Earth}]$}&\colhead{$[M_{\rm Jup}]$}
&\colhead{$[R_\odot]$} &\colhead{}  &\colhead{} 
\\
}
\startdata
9115800 & 4.454194338 & $12.21^{+7.19}_{-1.6}$ & n/a & $0.915^{+0.539}_{-0.12}$ & 6.5e-5 &KOI-421.01 
 \\
8544996 & 4.082275068 & $10.85^{+4.31}_{-0.81}$ & n/a & $0.707^{+0.316}_{-0.06}$& 2.1e-6 &KOI-913.01  \\
7832356 & 7.886631104 & $9.23^{+3.33}_{-1.62}$ & n/a & $1.135^{+0.409}_{-0.199}$ & 1.3e-11 &KOI-1456.01  \\
4076098&3.990106229&$8.02^{+3}_{-0.69}$&n/a&$0.9130^{+0.341}_{-0.079}$& 9e-5 &KOI1323.01  \\
9643874&8.027680595&$8.89^{+3.39}_{-0.63}$&n/a&$0.887^{+0.338}_{-0.063}$& 1.1e-3 &KOI1457.01  \\
7585481&8.098887986&$9.19^{+4.66}_{-0.85}$&n/a&$1.064^{+0.54}_{-0.098}$& 8.4e-9 &KOI890.01  
\\
3351888&1.625522200&$9.74^{+3.88}_{-1.24}$&n/a&$1.057^{+0.421}_{-0.135}$& 5.8e-7 &KOI801.01  
\\
9141746&6.491684259&$10.30^{+6.12}_{-1.24}$&n/a&$1.13^{+0.67}_{-0.14}$& 3.2e-4 &KOI-929.01 
\\
8255887&4.708326542&$11.10^{+4.17}_{-2.41}$&n/a&$1.21^{+0.46}_{-0.26}$& 4.9e-5 &KOI908.01 
\\
11138155&4.959319451&$11.88^{+4.91}_{-1.08}$&n/a&$1.025^{+0.424}_{-0.093}$& 1.7e-2 &KOI-760.01 
\\
10019708&3.268695154&$12.13^{+7.48}_{-1.4}$&n/a&$1.171^{+0.722}_{-0.135}$& 1.3e-10 &KOI-199.01  
\\
11414511&2.816504852&$12.82^{+4.94}_{-1.39}$&n/a&$0.9650^{+0.371}_{-0.105}$& 5.8e-4 &KOI-767.01  
\\
9595827&3.905081985&$13.14^{+0.86}_{-0.49}$&n/a&$0.8870^{+0.058}_{-0.033}$& 8.1e-4 &KOI-217.01 
 \\
12019440&3.243259796&$13.63^{+6.16}_{-1.23}$&n/a&$1.0290^{+0.465}_{-0.299}$& 1.7e-3 &KOI-186.01  
\\
8323764&6.714251076&$16.46^{+6.05}_{-1.79}$&n/a&$0.8^{+0.29}_{-0.084}$ &3.4e-2 & KOI-3767.01  
\\
6849046&4.225384512&$9.00^{+1.76}_{-0.66}$&n/a&$1.05^{+0.21}_{-0.08}$& 0 &KOI-201.01 \citepalias{Santerne:2015} 
\\
7778437&5.014234575&$10.31^{+7.57}_{-1.54}$&n/a&$1.31^{+0.26}_{-0.26}$& 0 & KOI131.01 \citepalias{Santerne:2012} 

\\
7017372&5.240904530&$12.84^{+6.69}_{-2.52}$&n/a&$1.41^{+0.59}_{-0.24}$& 0 &KOI-3689.01 \citepalias{Santerne:2015} 

\\
757450&8.884922680&$11.30^{+0.66}_{-0.66}$&$9.9^{+0.5}_{-0.5}$&$0.88^{+0.04}_{-0.04}$& 0 &Kepler-75b \citepalias{Bonomo:2015} 
\\
4570949&1.544928883&$14.92^{+1.32}_{-1.32}$&$2.01^{+0.37}_{-0.35}$&$1.32^{+0.08}_{-0.08}$& 0 & Kepler-76b \citepalias{Faigler:2013} 
\\
5357901&3.797018335&$10.73^{+0.24}_{-0.24}$&$0.25^{+0.08}_{-0.08}$&$0.86^{+0.02}_{-0.02}$& 0 &Kepler-425b \citepalias{He:14} 
\\
5358624&3.525632561&$11.85^{+0.33}_{-0.33}$&$1.27^{+0.19}_{-0.19}$&$0.80^{+0.02}_{-0.02}$& 0 & Kepler-428b \citepalias{He:14} 
\\
5728139&5.334083460&$15.91^{+1.76}_{-1.76}$&$2.82^{+0.52}_{-0.52}$&$2.26^{+0.25}_{-0.25}$& 0 &Kepler-433b \citepalias{Almenara:2015} 
\\
5780885&4.885488953&$17.78^{+0.11}_{-0.11}$&$0.44^{+0.04}_{-0.04}$&$1.96^{+0.07}_{-0.07}$& 0 & Kepler-7b \citepalias{Latham:2010} 
\\
6046540&7.340714746&$10.53^{+0.22}_{-0.22}$&$0.63^{+0.12}_{-0.12}$&$1.12^{+0.04}_{-0.04}$& 0 & Kepler-74b \citepalias{Bonomo:2015} 
\\
6922244&3.522498573&$15.58^{+0.55}_{-0.66}$&$0.59^{+0.13}_{-0.12}$& $1.45^{+0.12}_{-0.13}$& 0 & Kepler-8b \citepalias{Jenkins:2010} 
\\
7529266&8.600153301&$21.84^{+1.98}_{-1.98}$&$0.84^{+0.15}_{-0.15}$&$3.21^{+0.3}_{-0.3}$& 0 & Kepler-435b \citepalias{Almenara:2015} 
\\
7877496&1.720861324&$14.70^{+0.44}_{-0.55}$&$0.94^{+0.12}_{-0.02}$&$1.29^{+0.04}_{-0.04}$& 0 & Kepler-412b \citepalias{Deleuil:2014} 
\\
8191672&3.548465405&$15.69^{+0.44}_{-0.55}$&$2.11^{+0.07}_{-0.09}$&$1.75^{+0.14}_{-0.15}$&0 &Kepler-5b \citepalias{Koch:2010} 
\\
9305831&3.246732651&$11.96^{+0.77}_{-0.77}$&$1.00^{+0.1}_{-0.1}$&$1.35^{+0.08}_{-0.08}$&0 &Kepler-44b \citepalias{Bonomo:2012} 
\\
9410930&1.855557540&$11.41^{+0.44}_{-0.44}$&$0.56^{+0.10}_{-0.09}$&$1.02^{+0.03}_{-0.03}$&0 &Kepler-41b \citepalias{Esteves:2015} 
\\
9631995&7.891448474&$12.62^{+1.21}_{-1.21}$&$0.43^{+0.13}_{-0.13}$&$1.24^{+0.12}_{-0.12}$&0&Kepler-422b \citepalias{Endl:2014} 
\\
9651668&2.684328485&$13.17^{+0.77}_{-0.77}$&$0.72^{+0.12}_{-0.12}$&$0.99^{+0.05}_{-0.05}$&0&Kepler-423b \citepalias{Endl:2014} 
\\
9818381&3.024092548&$13.39^{+0.77}_{-0.66}$ &$3.23^{+0.26}_{-0.26}$ &$1.38^{+0.05}_{-0.03}$&0 & Kepler-43b \citepalias{Esteves:2015,Bonomo:2015} 
\\
10264660&6.790121599&$12.46^{+0.59}_{-0.59}$&$8.4^{+0.19}_{-0.19}$&$2.05^{+0.08}_{-0.08}$&0 & Kepler-14b \citepalias{Buchhave:2011} 
\\
10619192&1.485710952&$14.59^{+0.44}_{-0.44}$&$2.47^{+0.10}_{-0.10}$&$1.17^{+0.09}_{-0.09}$&0&Kepler-17b \citepalias{Bonomo:2012} 
\\
10666592&2.204735365&$15.69^{+1.2}_{-1.2}$&$1.741^{+0.028}_{-0.028}$&$2.00^{+0.01}_{-0.02}$&0&HAT-P-7b \citepalias{Pal:2008,Van:13} 
\\
10874614&3.234699312&$14.26^{+0.22}_{-0.33}$&$0.67^{+0.04}_{-0.04}$&$1.29^{+0.09}_{-0.10}$&0 &Kepler-6b \citepalias{Esteves:2015} 
\\
11017901&7.794301316&$18.10^{+6.47}_{-6.47}$&$1.37^{+0.48}_{-0.48}$&$1.03^{+0.16}_{-0.16}$&0 &Kepler-447b \citepalias{Lillo-box:2015} 
\\
11359879&4.942783399&$10.54^{+0.66}_{-0.77}$&$0.66^{+0.08}_{-0.09}$&$0.98^{+0.16}_{-0.06}$&0 &Kepler-15b \citepalias{Endl:2011} 
\\
11804465&4.437963030&$19.20^{+0.33}_{-0.44}$&$0.43^{+0.05}_{-0.05}$&$1.42^{+0.30}_{-0.24}$&0 &Kepler-12b \citepalias{Esteves:2015} 
\\
11446443&2.4706133738&$13.05^{+0.27}_{-0.27}$&$1.253^{+0.052}_{-0.052}$&$1.0^{+0.036}_{-0.036}$&0 &Tres-2b \citepalias{OD:06,Southworth:2010} 
\\
11502867&3.217518593&$11.96^{+0.33}_{-0.33}$&$0.34^{+0.08}_{-0.08}$&$0.92^{+0.02}_{-0.02}$&0 &Kepler-426b \citepalias{He:14} 
\\
9941662&1.763587569&$16.57^{+0.44}_{-0.44}$&$9.28^{+0.16}_{-0.16}$&$1.74^{+0.04}_{-0.04}$&0 &Kepler-13b \citepalias{Shporer:2011} 
\\
8359498&3.578780551&$10.53^{+0.22}_{-0.22}$&$0.43^{+0.03}_{-0.03}$&$0.99^{+0.02}_{-0.02}$& 0 &KOI127.01 \citepalias{Gandolfi:2013} 
\\

\enddata
\vspace{0.2cm}
\tablenotetext{a}{The false positive rate for individual system $fp$ is from the NASA Exoplanet Archive Kepler false positive probability table based on Morton et al (in prep).\footnote{http://exoplanetarchive.ipac.caltech.edu/cgi-bin/TblView/nph-tblView?app=ExoTbls$\&$config=koifpp} 
We assign 0 to the confirmed candidates. 
}
\tablenotetext{b}{{\y Reference abbreviations used in this table and
    the next are as follows.
\citetalias{Almenara:2015}:\citet{Almenara:2015};
\citetalias{Bonomo:2012}:\citet{Bonomo:2012};      
\citetalias{Bonomo:2015}:\citet{Bonomo:2015};        
\citetalias{Bruno:2015}:\citet{Bruno:2015};   
\citetalias{Buchhave:2011}:\citet{Buchhave:2011};
\citetalias{Cabrera:2013}:\citet{Cabrera:2013};    
\citetalias{Dawson:2012}:\citet{Dawson:2012};        
\citetalias{Deleuil:2014}:\citet{Deleuil:2014};    
\citetalias{Endl:2011}:\citet{Endl:2011};         
\citetalias{Endl:2014}:\citet{Endl:2014};  
\citetalias{Esteves:2015}:\citet{Esteves:2015};      
\citetalias{Faigler:2013}:\citet{Faigler:2013};     
\citetalias{Gandolfi:2013}:\citet{Gandolfi:2013};      
\citetalias{He:14}:\citet{He:14};
\citetalias{Jenkins:2010}:\citet{Jenkins:2010};
\citetalias{Koch:2010}:\citet{Koch:2010};     
\citetalias{Latham:2010}:\citet{Latham:2010};        
\citetalias{Lillo-box:2015}:\citet{Lillo-box:2015};     
\citetalias{Nesvorny:2012}:\citet{Nesvorny:2012};      
\citetalias{OD:06}:\citet{OD:06};
\citetalias{Pal:2008}:\citet{Pal:2008};           
\citetalias{Sanchis:2013}:\citet{Sanchis:2013};       
\citetalias{Santerne:2012}:\citet{Santerne:2012};      
\citetalias{Santerne:2015}:\citet{Santerne:2015};     
\citetalias{Schmitt:2014}:\citet{Schmitt:2014};   
\citetalias{Schmitt:2014b}:\citet{Schmitt:2014b};      
\citetalias{Shporer:2011}:\citet{Shporer:2011};     
\citetalias{Steffen:2010}:\citet{Steffen:2010};    
\citetalias{Southworth:2010}:\citet{Southworth:2010};    
\citetalias{Rowe:2014}:\citet{Rowe:2014};
\citetalias{Van:13}:\citet{Van:13};   
\citetalias{Weiss:2013}:\citet{Weiss:2013}.
}}
\end{deluxetable*}

\begin{deluxetable*}{llccccl}
\tabletypesize{\footnotesize}
\centering
\tablewidth{0pc}
\tablecaption{Kepler Warm Jupiter Systems \label{table:wjplanet}}

\tablehead{
\colhead{KIC}&\colhead{Period}&\colhead{Rp}&\colhead{Mp}&\colhead{Rstar}&
\colhead{fp rate} &\colhead{Other Name/Reference} \\ 
\colhead{}&\colhead{[day]}&\colhead{$[R_{\rm
    Earth}]$}&\colhead{$[M_{\rm
    Jup}]$}&\colhead{$[R_\odot]$}&\colhead{} &\colhead{} \\
}
\startdata
7984047&77.63425713&$10.09^{+1.35}_{-0.91}$&n/a&$0.755^{+0.101}_{-0.068}$& 8.8e-3 &KOI-1552.01
\\
7811397 &169.49954 &$9.89^{+2.26}_{-0.81}$ & n/a & $0.719^{+0.164}_{-0.059}$ & 0.71 &KOI-1477.01 \\
8672910&39.64317811&$10.52^{+4.46}_{-0.81}$&n/a&$0.833^{+0.355}_{-0.064}$& 3.6e-4 &KOI-918.01
\\
7504328&53.71797107&$8.21^{+2.73}_{-1.59}$&n/a&$1.151^{+0.383}_{-0.222}$& 0.46 &KOI-458.01 
\\
 6471021&125.62887621&$8^{+1.14}_{-0.34}$&n/a&$0.897 ^{+0.128}_{-0.038}$& 2.2e-4 &KOI-372.01 
\\
6061119&27.807562927&$14.46^{+6.63}_{- 1.08}$&n/a&$0.791^{+0.363}_{- 0.059}$& 7.7e-2 &KOI-846.01
\\
4760746&15.068059056&$17.65^{+5.88}_{-1.64}$&n/a&$0.96^{+0.32}_{-0.089}$& 0.8 &KOI-1455.01
 \\
10656508&124.03590546&$10.41^{+6.91}_{- 1.14}$&n/a&$1.137^{+0.754}_{-0.125}$&7.9e-3 &KOI-211.01  \\
7950644&10.290993755&$13.776\pm2.35$  & $0.29\pm0.09$ &$1.35\pm0.2$ &0&Kepler-427b \citepalias{He:14} \\
4164994&10.506825646&$9.41^{+1.89}_{-0.64}$&n/a&$0.784^{+0.157}_{-0.054}$&7.5e-4 & KOI-1320.01
\\
11194032&28.511205250&$8.21^{+6.23}_{-4.56}$&n/a&$1.741^{+1.322}_{-0.966}$& 0 & KOI-348.01  \\
7499398&23.020303180&$8.22^{+3.19}_{-0.68}$&n/a&$0.95^{+0.368}_{-0.079}$&9.4e-3 & KOI-1473.01  \\
7368664&12.874711418&$12.40^{+2.85}_{-1.98}$&$2.86^{+0.35}_{-0.35}$&$1.38^{+0.13}_{-0.13}$& 0 &Kepler-434b \citepalias{Almenara:2015} \\
7951018&52.75875577&$9.04^{+5.84}_{-1.04}$&n/a&$1.154 ^{+0.746}_{-0.132}$& 8.5e-4 &KOI-1553.01 
\\
9025971&141.24164672&$10.64^{+4.02}_{-0.85}$&$0.55 ^{+0.02}_{-0.02}$&$0.913^{+0.345}_{- 0.073}$&0&KOI-3680.01 \citepalias{Santerne:2015}
\\
 5812701&17.855219698&$15.69^{+1.43}_{-1.43}$&n/a&$1.63^{+0.15}_{-0.15}$& 0 &KOI-12.01 \citepalias{Bonomo:2015} \\
10723750&50.79034619&$9.16^{+5.55}_{-1.14}$&$1.84^{+ 0.18}_{-0.183}$&$1.183^{+0.716}_{- 0.147}$&0 &Kepler-117c \citepalias{Rowe:2014,Bruno:2015} \\
 &18.795900480&$6.04^{+3.66}_{-0.75}$&$0.094 ^{+0.03}_{-0.03}$& & 0&Kepler-117b \citepalias{Rowe:2014,Bruno:2015} \\
7109675&33.601220660&$8.83^{+0.46}_{-0.5}$&n/a&$0.93^{+0.048}_{-0.053}$& 0 &Kepler-46b \citepalias{Nesvorny:2012} 
\\
&57.011&n/a&$0.376^{+0.021}_{-0.019}$&
 &0&Kepler-46c \citepalias{Nesvorny:2012}  
\\
&6.76652078&$2.17^{+0.11}_{-0.13}$&n/a&
&0&Kepler-46d \citepalias{Nesvorny:2012}  
\\
6462863&22.342969585&$9.77^{+2.03}_{-1.5}$&$0.333^{+ 0.036}_{-0.036}$ & $1.297^{+0.27}_{-0.199}$ &0&Kepler-89d \citepalias{Weiss:2013}   \\
&10.423677765&$3.66^{+0.76}_{-0.56}$&$0.030\pm 0.015$&
&0&Kepler-89c \citepalias{Weiss:2013}  
\\
&54.31998605&$5.79^{+1.21}_{-0.89}$&$0.1101\pm0.045 $&
&0&Kepler-89e \citepalias{Weiss:2013} 
\\
&3.743175556&$1.55^{+0.32}_{-0.24}$&$0.02\pm0.02$&
&0&Kepler-89b \citepalias{Weiss:2013} 
\\
5972334&15.358768403&$11.31^{+4.78}_{-1.04}$& n/a&$0.87^{+0.368}_{-0.08}$&4.1e-4 &KOI-191.01 \citepalias{Steffen:2010} \\
&2.418405445&$2.79^{+1.17}_{-0.26}$& n/a&
& 3e-3 &KOI-191.02  \citepalias{Steffen:2010}  \\
&0.708620008 & $1.25^{+0.52}_{-0.12}$&n/a&  
& 1.0 &KOI-191.03 \citepalias{Steffen:2010}  \\
&38.6519976&$2.62^{+1.11}_{-0.24}$&n/a&
& 1.3e-3 
&KOI-191.04 \citepalias{Steffen:2010}  \\
5629353&33.319916700&$13.10^{+7.22}_{-3.66}$&n/a& $1.63^{+0.897}_{-0.456}$& 2e-2 &KOI-6132.01 
\\
&7.5844126&$5.49^{+3.03}_{-1.53}$& n/a&
& 3.8e-4&KOI-6132.02
\\
&11.8674823&$3.94^{+2.17}_{-1.1}$&n/a& 
& 3.7e-3 &KOI-6132.03
\\
9946525&51.84688575&$8.95^{+2.07}_{-0.65}$&n/a&$0.838 ^{+0.252}_{-0.061}$& 4.43-3 &KOI-398.01
\\
&4.18004955&$3.55^{+1.07}_{-0.26}$&n/a&
& 3.8e-7 &Kepler-148c 
 \\
&1.729366467&$1.87^{+0.56}_{-0.15}$&n/a& 
& 0.88 &Kepler-148b  
\\
3832474&143.2063518&$9.36^{+0.99}_{-0.37}$& $0.073 ^{+0.008}_{-0.008}$&$0.867^{+0.092}_{- 0.034}$& 0&Kepler-30d \citepalias{Sanchis:2013}  \\
&60.32488611&$12.88^{+1.36}_{-0.51}$&$2.01 ^{+0.16}_{-0.16}$& 
&0&Kepler-30c \citepalias{Sanchis:2013} 
\\
&29.1598615&$1.91^{+0.2}_{-0.07}$&$0.036^{+0.004}_{-0.004}$&
&0&Kepler-30b \citepalias{Sanchis:2013}
 
\\
3247268&86.67855186&$9.88^{+4.67}_{-1.02}$&$1.1^{+1.1}_{-0.0}$&$1.043^{+0.494}_{-0.107}$&0&Kepler-418b\\
&12.218278&$4.75^{+2.25}_{-0.49}$&n/a& 
& 2.0e-4  &KOI-1089.02 
\\
7303287&125.8518&$11.35^{+0.19}_{-0.19}$&$0.41^{+0.05}_{-0.05}$
&$1.00^{+0.02}_{-0.02}$& 0 &Kepler-289b \citepalias{Schmitt:2014b}\\ 
&34.5438464& $2.11^{+0.1}_{-0.1}$&n/a&
& 0&Kepler-289c \citepalias{Schmitt:2014b}  
\\
&66.063& $2.62^{+0.16}_{-0.16}$&
$0.013^{+0.003}_{-0.003}$&
&0&PH-3c \citepalias{Schmitt:2014b}  
\\
12365184&69.7546&$10.6^{+1.4}_{-1.3}$& $2.5^{+0.3}_{- 0.3}$&$1.75^{+0.08}_{-0.07}$&0&Kepler-419b \citepalias{Dawson:2012} \\
&675.47&n/a&$7.3^{+0.4}_{-0.4}$&
&0&Kepler-419c \citepalias{Dawson:2012} 
\\
7898352&127.2824031&$9.32^{+3.76}_{-0.77}$&n/a& $0.883^{+0.356}_{-0.073}$& 7.2e-2 &Kepler-302c \citepalias{Rowe:2014} \\ 
&30.1836854&$2.88^{+1.15}_{-0.24}$&n/a&
& 3.4e-5 &Kepler-302b \citepalias{Rowe:2014}
 
\\

\enddata
\vspace{0.2cm}
\label{tab:table2}
\end{deluxetable*}



\bibliographystyle{apj_eprint}





\end{document}